\input{aipcheck}

\documentclass[draft]{aipproc}

\layoutstyle{6x9}

\begin{document}

\title{Concentrating Energy by Measurement}


\author{A. Beige}{
  address={The School of Physics and Astronomy, University of Leeds,
  Leeds, LS2 9JT, United Kingdom}
}

\author{A. Capolupo}{
  address={Dipartimento di Matematica e Informatica dell'Universit\`a di Salerno and\\
 Istituto Nazionale di Fisica Nucleare, Gruppo Collegato di Salerno, 84100
 Salerno, Italy}
}

\author{E. Del Giudice}{
  address={I.N.F.N. Sezione di Milano, Universit\`a di Milano, I-20133 Milano, Italy}
}

\author{A. Kurcz}{
  address={The School of Physics and Astronomy, University of Leeds,
  Leeds, LS2 9JT, United Kingdom}
}

\author{G. Vitiello}{
  address={Dipartimento di Matematica e Informatica dell'Universit\`a di Salerno and\\
 Istituto Nazionale di Fisica Nucleare, Gruppo Collegato di Salerno, 84100
 Salerno, Italy}
}

\begin{abstract}
In a recent article [A. Kurcz {\em et al.}, Phys. Rev. A {\bf 81}, 063821 (2010)] we predicted an energy concentrating mechanism in composite quantum systems. Its result is a non-zero stationary state photon emission rate even in the absence of external driving. Here we discuss the possible origin of the predicted effect. We attribute it to the presence of a non-trivial interaction between different system components and to repeated environment-induced photon measurements.
\end{abstract}

\maketitle

\section{Introduction}

Already in 1936, Dirac asked the question, whether conservation of energy holds in atomic processes \cite{Dirac}. He commented on a theory of radiation which had been put forward by Bohr, Kramers, and Slater  \cite{Bohr} in 1924 and which violated the conservation of energy. Their theory gave no conservation of energy for individual atomic processes, though it gave statistical conservation of energy when large numbers of atomic processes take place. Dirac argued in favor of some part of this theory and suggested that energy might indeed not be preserved in processes involving large velocities, including radiative processes. However, he also pointed out that there is a primitive theory of radiation which gives information about the probability of a {\em quantum jump} under the influence of external radiation which is equally consistent with the conservation laws and with the main assumptions of the radiation theory by Bohr {\em et al.} \cite{Bohr}.

Another discussion on the conservation of energy in atomic processes can be found by Milonni in his book on the quantum vacuum \cite{Milonni}. In Chapter 2.6, he considers a linear dipole oscillator in the vacuum. Starting from the minimal coupling Hamiltonian for the interaction between the particle and the surrounding free radiation field, he calculates the time evolution of the position of the particle. Although he considers a dipole in the vacuum without any external field acting on it, he observes a non-trivial time evolution. Its origin is the effect of the vacuum field on the dipole \cite{Milonni}. Today we know that Milonni's predictions cannot be verified easily experimentally. The reason is that the so-called rotating wave approximation which makes the effect of the vacuum onto a dipole disappear works in general very well in the situation considered by Milonni. In this paper, we discuss a closely related energy concentrating effect in {\em composite quantum systems} which should be observable experimentally, even with currently available technology.

In classical physics, coupled systems prepared in their respective lowest-possible energy state cannot exchange energy. However, this statement does not necessary apply to quantum systems. The energy ground state of a composite quantum system is in general not the product of the energy ground states of its parts. Due to interactions between different components, the energy ground state of a composite quantum system is in general an entangled state. The energy expectation value for the product of the energy ground states of the individual systems is hence in general higher than the ground state energy of the system. In such a case, a composite quantum system with each of its components initially prepared in its respective energy ground state evolves in time, even in the absence of external driving. If a measurement is performed, for example, on one of the system components whether it is in its ground state or not, there is a non-zero probability to find this component in an excited energy eigenstate. When this happens, it might look as if energy appears from nowhere \cite{Andreas,Schulman06}.

\begin{figure} \label{fig1}
  \includegraphics[height=.3 \textheight]{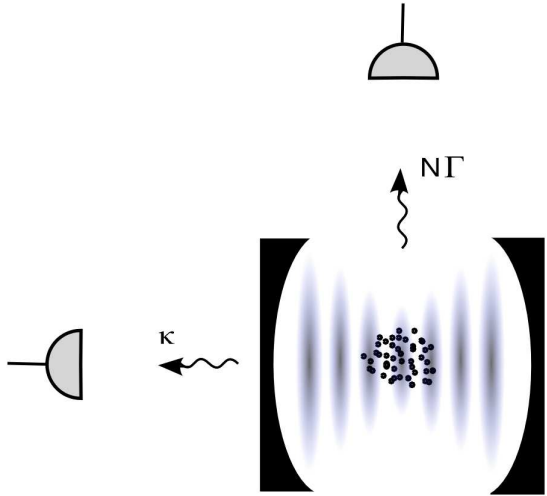} \\[0.4cm]
  \caption{Experimental setup showing an large number of $N$ atoms trapped inside an optical cavity. The atoms should be well localised within an optical coherence domain so that their behaviour becomes collective. The spontaneous decay rate of atomic excitation is then given by $N \Gamma$. The optical resonator couples separately to an orthogonal set of modes of the free radiation field with $\kappa$ denoting the corresponding spontaneous decay rate.}
\end{figure}

In this paper, we discuss a concrete example of the above described energy-concentrating mechanism in composite quantum systems. As illustrated in Figure \ref{fig1}, we consider a large number of atoms trapped inside an optical cavity. We then show that the energy ground state of the atom-cavity system differs indeed from the product of the energy ground state of an ensemble of non-interacting atoms and the energy ground state of an optical cavity field. The reason for this are the normally-neglected counter-rotating terms in the atom-cavity interaction Hamiltonian. Moreover, we point out that the environment probes the energy of the individual components of this composite quantum system. The reason is that atoms and cavity couple to different parts of the surrounding free radiation field. It is argued that detecting either the atoms or the cavity in their respective energy ground states pumps energy in the system. Repeated environment-induced photon measurements hence manifest themselves in the continuous leakage of photons through the cavity mirrors, even in the absence of external driving.\footnote{From our calculations below we see that these measurements are not frequent enough to suppress the internal system dynamics, as predicted by the Quantum Zeno effect \cite{Misra}.}

The energy-concentrating mechanism discussed in this paper with a single atom inside an optical cavity was first pointed out by Werlang {\em et al.} \cite{Werlang08} in 2008. In a recent paper \cite{Kurcz10}, we analyzed the predicted effect in much more detail. Considering a collectively enhanced version of the energy concentration in atom-cavity systems based on the trapping of not only one but many atoms inside the resonator, we predicted stationary state photon emission rates whose observation is in principle feasible with current technology. The reason for this is the recent development of atom-cavity systems mounted on atom chips with relatively large atom-cavity coupling constants (due to very small cavity mode volumes) and relatively large spontaneous cavity decay rates \cite{Trupke}. Experimental evidence for the presence of the counter-rotating terms in atom-cavity interaction Hamiltonians in the form of level shifts has recently been observed in systems which combine a superconducting qubit with a microwave resonator \cite{Mooji}. Such systems are hence another promising candidate for the experimental verification of the possibility of concentrating energy by measurement \cite{Sabin10}.

As already pointed out above, we attribute the energy concentration in atom-cavity systems to the counter-rotating terms in the atom-cavity interaction Hamiltonian as well as to the effect of the environment onto the surrounding free radiation field. The detection of a single photon can indeed spontaneously increase or decrease the energy of a {\em single} quantum system. To illustrate this, we now consider a single two-level atom with ground state $|1 \rangle$ and excited state $|2 \rangle$. Suppose the atom is initially in a superposition and its state can be written as
\begin{eqnarray}
|\psi \rangle &=& \alpha \, |1 \rangle + \beta \, |2 \rangle
\end{eqnarray}
with $|\beta|^2 < 1$. Denoting the energy of the excited state $|2 \rangle$ by $\hbar \omega_{\rm a}$ and the energy of the ground state by zero, one can easily see that the energy stored in the atom is given by $|\beta|^2 \, \hbar \omega_{\rm a}$. Whenever the atom spontaneously emits a photon, i.e.~when a quantum jump occurs, the atom releases in general a larger amount of energy into the environment, since the energy of the emitted photon equals $\hbar \omega_{\rm a}$. However, there is also the possibility that the atom does not emit a photon. In this case, the atom releases no energy. Instead it undergoes a non-radiative transition into its ground state $|1 \rangle$ \cite{Hegerfeldt93}. This spontaneous creation of energy is an effect of the quantum vacuum {\em and} of the environment which detects the photon.

There are five sections in this paper. In the next Section we introduce the theoretical model used throughout this paper. We then summarize the results of Ref.~\cite{Kurcz10} and emphasize the role of the counter-rotating terms in the system dynamics. Afterwards, we emphasize the role of environment-induced photon measurements for the above mentioned energy concentrating mechanism. Finally, we summarize our results and point out related effects, like the Casimir effect and the well-known existence of Lamb-shifts, in other quantum systems.

\section{Theoretical model}

Let us now have a closer look at the experimental setup shown in Figure \ref{fig1} which contains a large number $N$ of identical tightly confined atoms inside an optical cavity. The energy of this composite system is the sum of the free energy of both subsystems, their interactions with the surrounding free radiation field, and the interaction between the atoms and the cavity field. The Hamiltonian $H$ of the system in dipole approximation and in the Schr\"odinger picture can hence be written as
\begin{eqnarray} \label{H2}
H &=& H_0 + H_{\rm int}
\end{eqnarray}
with
\begin{eqnarray} \label{H3}
H_0 &=& \hbar ~ \omega_{\rm c} ~c^\dagger c + \hbar \omega_{\rm a} ~ S^+ S^-  + \sum _{{\bf k} \lambda} \hbar \omega _k ~ a ^\dagger _{{\bf k} \lambda} a _{{\bf k} \lambda} ~ , \nonumber \\
H_{\rm int} &=& \sum_{{\bf k} \lambda} \hbar \big( g_{{\bf k} \lambda} ~ a _{{\bf k} \lambda} + \tilde g_{{\bf k} \lambda} ~ a _{{\bf k} \lambda}^\dagger \big) \big( c + c^\dagger \big) + \sqrt{N} \hbar ~ \big( q_{{\bf k} \lambda} ~ a _{{\bf k} \lambda} + \tilde q_{{\bf k} \lambda} ~ a _{{\bf k} \lambda}^\dagger \big) \big( S^+ + S^- \big) \nonumber \\
&& + \sqrt{N} \hbar g_{\rm c} ~ \big( c+c^\dagger \big) \big( S^+ + S^- \big) \, .
\end{eqnarray}
Here $S^-$, $c$, and $a_{{\bf k} \lambda}$ are bosonic annihilation operators for collective atomic excitations, cavity photons and photonic excitations of the mode $({\bf k},\lambda)$ of the free radiation field, respectively. The variables $g_{\rm c}$, $g_{{\bf k} \lambda}$, $\tilde g_{{\bf k} \lambda}$, $q_{{\bf k} \lambda}$ and $\tilde q_{{\bf k} \lambda}$ denote coupling constants and $\omega_{\rm c}$, $\omega_{\rm a}$, and $\omega_k$ are frequencies. The vector ${\bf k}$ is as usual a wave vector and $\lambda$ is a polarization. Here the cavity photon states have been chosen such that $g_{\rm c}$ is automatically real. The atoms should be well localized within an optical coherence domain. Such high mode densities lead to a strong coupling regime which is responsible for the collective enhancement factor $\sqrt{N}$ in front of atomic interaction terms \cite{Holstein,Shah,Beige}.

Proceeding as usual \cite{Hegerfeldt93,Dalibard92,Hegerfeldt94}, the Hamiltonian in Eq.~(\ref{H2}) can be used to predict the time evolution of the density matrix $\rho$ of the atom-cavity system alone in the presence of non-zero spontaneous decay rates. It is given by the master equation of the form \cite{Kurcz10}
\begin{eqnarray} \label{deltarho2}
\dot \rho &=& - {{\rm i} \over \hbar} \left[ H_{\rm cond} \rho - \rho H_{\rm cond}^\dagger \right] + {\cal R}(\rho)
\end{eqnarray}
with
\begin{eqnarray} \label{last2}
H_{\rm cond} &=& \hbar \Big( \widetilde \omega_{\rm c} - {{\rm i} \over 2} \kappa \Big) ~ c^\dagger c + \hbar \Big( \widetilde \omega_{\rm a} - {{\rm i} \over 2} N \Gamma \Big) ~ S^+ S^- + \sqrt{N} \hbar g_{\rm c} ~ \big( c+ c^\dagger \big) \big( S^+ + S^- \big) ~, \nonumber \\
{\cal R}(\rho) &=& \kappa~ c \rho c^\dagger + N\Gamma ~ S^+ \rho S^- ~.
\end{eqnarray}
Here $\widetilde \omega_{\rm c}$ and $\widetilde \omega_{\rm a}$ denote the bare atom and cavity frequencies, $\kappa$ is the cavity decay rate, and $\Gamma$ is the decay rate of the excited state of a single atom. In the derivation of this equation, we neglected the counter-rotating terms in the system-bath interaction in Eq.~(\ref{H3}). These are known to lead in general only to very small effects and neglecting them is in good agreement with actual quantum optics experiments.

In the following, we are especially interested in the mean number of photons inside the optical cavity. To analyze the dynamics induced by the above master equation we therefore consider the expectation values
\begin{eqnarray}
&& \mu_1 \equiv \langle c^\dagger c \rangle \, ~~ \mu_2 \equiv \langle S^+ S^- \rangle \, , ~~ \eta_{1,2} \equiv {\rm i} \langle (S ^- \pm S ^+) (c \mp c^\dagger) \rangle\, , ~~\nonumber \\
&& \eta _{3,4} \equiv\langle (S ^- \mp S ^+) (c \mp c^\dagger )\rangle\, , ~~\xi_1 \equiv {\rm i} \langle c^2 - c^{\dagger 2} \rangle\, , ~~\xi_2 \equiv \langle c^2 + c^{\dagger 2} \rangle\, , ~~ \nonumber \\
&& \xi_3 \equiv {\rm i} \langle S^{-2} - S^{+2} \rangle\, , ~~\xi_4 \equiv \langle S^{-2} + S^{+2} \rangle \, .
\end{eqnarray}
Using Eq.~(\ref{deltarho2}), one can show \cite{Andreas,Kurcz10} that the time evolution of these variables is given by a closed set of rate equations. These are
\begin{eqnarray} \label{rates}
&&\dot \mu _1 = \sqrt{N} g_{\rm c} \eta _1 - \kappa \mu _1 \, , ~~
\dot \mu _2 = \sqrt{N}  g_{\rm c} \eta _2 - N \Gamma \mu _2 ~ , \nonumber \\
&&\dot \eta _1 =  2 \sqrt{N} g_{\rm c} (1 + 2 \mu_2 + \xi_4) + \widetilde \omega_{\rm a} \eta _3 + \widetilde \omega_{\rm c} \eta _4 - {\textstyle {1 \over 2}} \zeta \eta_1 ~ , \nonumber \\
&&\dot \eta _2 = 2 \sqrt{N}  g_{\rm c} (1 + 2 \mu_1 + \xi_2) + \widetilde \omega _{\rm a} \eta _4 + \widetilde \omega _{\rm c} \eta _3 - {\textstyle {1 \over 2}} \zeta \eta_2 ~ , \nonumber \\
&&\dot \eta _3 = - 2 \sqrt{N}  g_{\rm c} (\xi_1 + \xi_3) - \widetilde \omega _{\rm a} \eta _1 - \widetilde \omega _{\rm c} \eta _2 - {\textstyle {1 \over 2}} \zeta \eta _3 ~ , \nonumber \\
&&\dot \eta _4 = - \widetilde \omega _{\rm a} \eta _2 - \widetilde \omega_{\rm c} \eta _1 - {\textstyle {1 \over 2}} \zeta \eta_4 ~, ~~ \dot \xi _1 = 2 \sqrt{N} g_{\rm c} \eta_4 + 2 \widetilde \omega_{\rm c} \xi_2 - \kappa \xi_1 ~, \nonumber \\
&&\dot \xi _2 = - 2 \sqrt{N} g_{\rm c} \eta_1 - 2 \widetilde \omega_{\rm c} \xi_1 - \kappa \xi_2 ~, ~~
\dot \xi _3 = 2 \sqrt{N} g_{\rm c} \eta_4 + 2 \widetilde \omega_{\rm a} \xi_4 - N \Gamma \xi_3 ~, \nonumber \\
&&\dot \xi _4 = - 2 \sqrt{N} g_{\rm c} \eta_2 - 2 \widetilde \omega_{\rm a} \xi_3 - N \Gamma \xi_4
\end{eqnarray}
with $\zeta \equiv \kappa + N \Gamma$. Calculating the stationary state photon emission rate of the atom-cavity system can now be done by setting all of the above time derivatives equal to zero and calculating the stationary state of the above rate equations.

\section{The atom-cavity interaction}

Let us now have a closer look at the atom-cavity interaction Hamiltonian. The crucial difference to the usual Jaynes-Cummings model \cite{Knight} is the presence of the $c S^-$ and the $c^\dagger S^+$ term in (\ref{H3}) which vanish in the rotating wave approximation (RWA). In the RWA, the ground state of the atom-cavity system would simply be given by $|E_0 \rangle = |0_{\rm a} \rangle |0_{\rm c} \rangle$ which is the product of the zero-excitation state $|0_{\rm a} \rangle$ of the atoms and the vacuum state $|0_{\rm c} \rangle$ of the cavity field. However, taking the counter-rotating terms into account, we find instead that the energy ground state of the atom-cavity systems is of the form
\begin{eqnarray}
|E_0 \rangle &=& |0_{\rm a} \rangle |0_{\rm c} \rangle + {\cal O}\left( {\sqrt{N} g_{\rm c} \over \omega_{\rm c}} \right)
\end{eqnarray}
when atoms and cavity are on resonance, i.e.~when $\omega_{\rm c} = \omega_{\rm a}$. This is an entangled state \cite{Capolupo10}. Let us now have a closer look at what this implies.

\begin{figure} \label{fig2}
  \includegraphics[height=.4\textheight]{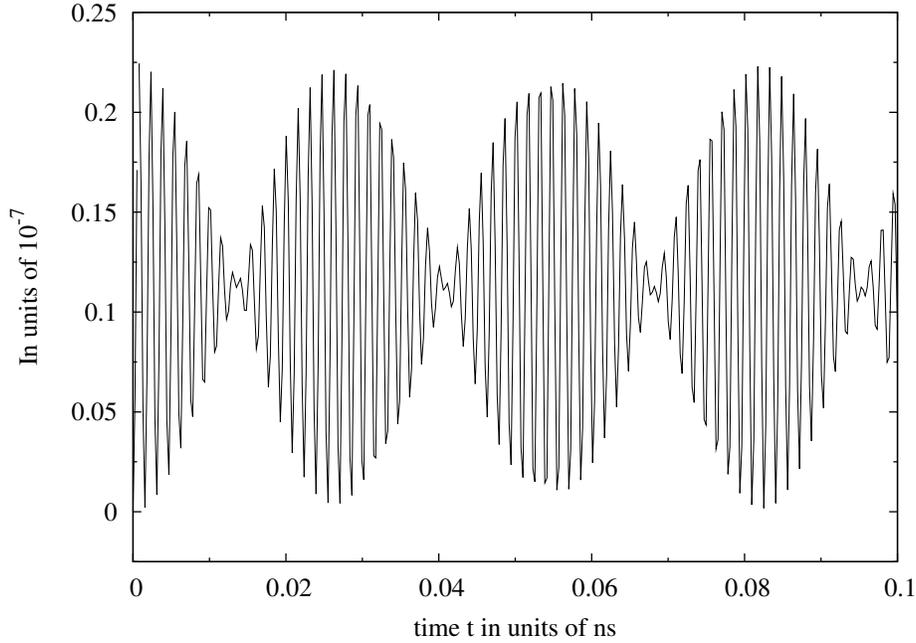}
  \caption{Time evolution of the mean number of photons in the cavity obtained from a numerical solution of the rate equations in Eq.~(\ref{rates}) for $\tilde \omega_{\rm c} = \tilde \omega_{\rm a} = 384.2 \cdot 10^{12} \, {\rm s}^{-1}$ (D2 line), $g_{\rm c} = 6.1 \cdot 10^8 \, {\rm s}^{-1}$, $\Gamma = 1.9 \cdot 10^7 \, {\rm s}^{-1}$, and $\kappa = 1.3 \cdot 10^{10} \, {\rm s}^{-1}$. This means, the experimental parameters considered here are the same as in Ref.~\cite{Trupke}. The atoms and the cavity were initially both in their respective lowest energy eigenstate $|0_{\rm a} \rangle$ and $|0_{\rm a} \rangle$, respectively. On average there is always a non-negligible amount of population inside the cavity field. Although being small, this population can result in a relatively large stationary state photon emission rate when multiplied with a sufficiently large spontaneous cavity decay rate $\kappa$.}
\end{figure}

Suppose, the atom-cavity system is initially prepared in $|0_{\rm a} \rangle |0_{\rm c} \rangle$. This state is one of the possible outcome of an environment-induced measurement on the atoms {\em and} on the cavity, whether each of these two system components is in an excited state or not. Since this measurement does not project the composite quantum system into one of its energy eigenstates, its state evolves in time due to the counter-rotating terms in the atom-cavity interaction Hamiltonian. These include a term proportional to $S^+ c^\dagger $ and simultaneously create atomic and photonic excitations. This behavior of atom-cavity systems is illustrated in Fig.~\ref{fig2} which shows the time evolution of the mean cavity photon number $\mu_1$. The average cavity photon number is of the order of $N g_{\rm c}^2 / \omega_{\rm c}^2 $. Its result is the spontaneous emission of photons even in the absence of external driving.

Using the rate equations in Eq.~(\ref{rates}) and calculating the stationary state cavity photon emission rate we find that it equals \cite{Kurcz10}
\begin{eqnarray} \label{IN}
I_{\kappa} &=& {N \zeta \kappa g_{\rm c}^2 \left [ \, 8 \zeta g_{\rm c}^2 + \zeta^2 \Gamma + 4 \Gamma \left ( \widetilde \omega _{\rm a} - \widetilde \omega_{\rm c} \right )^2 \, \right] \over 16 \zeta^2 g_{\rm c}^2 \widetilde \omega_{\rm a} \widetilde \omega_{\rm c} + 2 \zeta^2 \kappa \Gamma \left ( \widetilde \omega _{\rm a} ^2 + \widetilde \omega_{\rm c}^2 \right ) + 4 \kappa \Gamma \left( \widetilde \omega _{\rm a}^2 - \widetilde \omega_{\rm c}^2 \right)^2}
\end{eqnarray}
which applies for $N \Gamma, \sqrt{N} g_{\rm c}, \kappa \ll \widetilde \omega_{\rm a},\widetilde \omega_{\rm c}$. For example, the parameters of the recent cavity experiment with $^{85}$Rb $[4]$ combined with $N=10^4$ are expected to result in an $I_\kappa$ as large as $I_\kappa =301~\rm{s}^{-1}$. This rate is of the order of typical detector dark count rates. The verification of the above predicted signal should therefore be feasible with current technology. Important is to notice here that $I_\kappa$ is a function of the atom-cavity system parameters. Measuring such a system parameter dependence should make it possible to distinguish background photons from photons which have been generated through energy concentration by measurement.

\section{The role of the environment}

A crucial role in the energy concentration in atom-cavity system in Fig.~\ref{fig1} is played by the environment. In fact, the derivation of the master equation in Eq.~(\ref{deltarho2}) requires the assumption of a non-trivial coupling of the system to a surrounding free radiation field. It also requires the assumption of a photon-absorbing environment. As in Refs.~\cite{Hegerfeldt93,Dalibard92,Hegerfeldt94}, we consider in the following a coarse grained time scale $\Delta t$. On this time scale, the free radiation field is constantly re-set into its vacuum state $|0_{\rm ph} \rangle$. Immediately after such a resetting, the total density matrix $\rho$ of the system and the radiation field can be written as
\begin{eqnarray} \label{rho}
\rho &=& |0_{\rm ph} \rangle \, \rho_{\rm S} \, \langle 0_{\rm ph} | \, ,
\end{eqnarray}
where $\rho_{\rm S}$ is the state of the atom-cavity system. Assuming a resetting of the free radiation field onto the zero-photon state requires the absence of thermal photons in the free radiation field. One can check that this applies for photons in the optical regime even at relatively high temperatures, like room temperature.

Between two resettings, the atom-cavity system and the surrounding free radiation field evolve with the Hamiltonian given in Eq.~(\ref{H3}). One can easily see that the system-bath interaction terms in the Hamiltonian transfer energy from the atom-cavity system into the free radiation field, whenever either the atoms or the cavity are excited. The overall effect of this time evolution and the constant resetting of the free radiation field on a coarse grained time scale $\Delta t$ is the same as performing repeated measurements whether the atoms and the cavity mode are in their respective lowest energy eigenstates $|0_{\rm a} \rangle$ and $|0_{\rm c} \rangle$, respectively \cite{Hegerfeldt93}. The reason is that the atoms and the optical cavity couple separately to the different modes of the free radiation field. This means, the environment performs repeated energy-measurements on each component of the composite quantum system considered here. It is therefore not surprising that the dynamics of the system could be as described in the previous section.

\section{Conclusions}

From a general perspective a phenomenon like the energy concentration in a composite quantum system can indeed be motivated physically. There exist processes, where there is a redistribution of energy
among different system degrees of freedom making possible some amounts of system self-organization. In particular, one could examine the possibility of concentrating the total energy of the system into a subset of degrees of freedom producing a decrease of its entropy, which in order to avoid a violation of the second law of thermodynamics, would compel the release of energy to the environment, thus keeping the free energy constant. This is possible only if the system is {\it open}.

As predicted in Refs.~\cite{Kurcz10,Capolupo10}, a leakage of energy from a quantum system can occur among different degrees of freedom. This leakage is not necessarily triggered by an external pump of energy, but could be also triggered by virtual photons coming from the quantum vacuum as, e.g., it occurs in the Casimir effect or in the Lamb shift \cite{Milonni}. From the standpoint of the receiving system, the origin of the triggering energy is not important as far as the balance between the variations of energy and entropy is satisfied so to keep the free energy constant. In this respect, it is to recall that the ratio between these
variations is just the temperature, as required by the thermodynamic definition
\begin{eqnarray}
k_{\rm B} T &=& {{\mathrm d}U \over {\mathrm d}S} ~.
\end{eqnarray}
It is hence extremely appealing to study a dynamics, where a system is able to reach a state having a lower energy jumping over a separating barrier with the help of excitations from the vacuum. This will only become amenable, if the open system dynamics is {\it irreversible}.

Indeed, in this context Lamb shift and Casimir effect \cite{Milonni}, respectively, are widely understood observations that show how non-trivial zero-temperature properties of the system can arise due to the vacuum. However, in both of these examples the vacuum plays an immanent role in terms of renormalizing the original parameters of the system. Instead, the energy fluctuations which we focus on here cannot be obtained from the uncoupled system simply by renormalizing its parameters. As mentioned before, a closed system in its ground state does not fluctuate!

In this paper, we examined the effect of the counter-rotating terms in the context of open quantum systems. The mathematical analysis we have done shows that in an atom-cavity system decoherence can occur among different degrees of freedom. The subsequent leakage of photons is not necessarily triggered by an external pump of energy, but could also be triggered by virtual photons coming from the quantum vacuum and the presence of a photon detecting environment. Indeed, the interplay between the microscopic quantum dynamics and the thermal properties of a system are extremely appealing. Concerning such an interplay, we observe that quantum optical systems have the potential to simulate a variety of fundamental quantum effects, including for example Hawking radiation and the Unruh effect, which would otherwise not be as easily accessible experimentally \cite{Ford,Ford2,Cirac}.

\begin{theacknowledgments}
A.~B. and A.~K. would like to thank K.~M{\o}lmer, A.~Stokes, and T.~P.~Spiller for stimulating discussions. A.~B. acknowledges a James Ellis University Research Fellowship from the Royal Society and the GCHQ. This work was moreover supported by the UK Research Council EPSRC, the University of Salerno, and INFN.
\end{theacknowledgments}

\bibliographystyle{aipproc}

\end{document}